\documentclass[%
 aip,
 jmp,%
 amsmath,amssymb,
 preprint,%
]{revtex4-1}

\usepackage{graphicx}
\usepackage{dcolumn}
\usepackage{bm}

\newcommand\ot{\otimes}

\renewcommand\ker{{\text{\rm Ker}}\,}

\newcommand\inte{{\text{\rm int}}\,}
\newcommand\ttt{{\text{\rm t}}}
\newcommand\spa{{\text{\rm span}}\,}

\newcommand\rk{{\text{\rm rank}}\,}

\begin{document}

\preprint{AIP/123-QED}

\title
{Classification of bi-qutrit positive partial transpose entangled edge states by their ranks}

\author{Seung-Hyeok Kye}
  \affiliation
 {Department of Mathematics and Institute of Mathematics,
 Seoul National University, Seoul 151-742, Korea}

\author{Hiroyuki Osaka}
\affiliation{%
Department of Mathematical Sciences,   Ritsumeikan University, Kusatsu, Shiga, 525-8577  Japan
}%

\date{\today}

\begin{abstract}
We construct $3\otimes 3$ PPT entangled edge states with maximal
ranks, to complete the classification of $3\otimes 3$ PPT entangled
edge states by their types. The ranks of the states and their
partial transposes are $8$ and $6$, respectively. These examples
also disprove claims in the literature.
\end{abstract}

\pacs{03.67.-a, 03.67.Hk, 03.65.Fd}
\keywords{positive partial transposes, separable states,
entanglement, edge states, product vectors}
\maketitle

\section{Introduction}

Let $M_n$ denote the $C^*$-algebra of all $n\times n$ matrices over
the complex field, with the cone $M_n^+$ of all positive
semi-definite matrices. A positive semi-definite matrix in
$M_m\otimes M_n$ is said to be separable if it is the convex sum of
rank one projectors onto product vectors $x\ot y\in\mathbb
C^m\ot\mathbb C^n$. A positive semi-definite matrix in $M_m\otimes
M_n$ is said to be entangled if it is not separable. Since the
convex cone of all separable ones coincides with $M_m^+\otimes
M_n^+$, the entanglement consists of $(M_m\otimes M_n)^+\setminus
M_m^+\otimes M_n^+$. The notion of entanglement is a unique
phenomenon in non-commutative order structures, and there is no
counterpart in classical mechanics. Indeed, it is well-known that
the equality $({\mathcal A}\otimes {\mathcal B})^+ ={\mathcal
A}^+\otimes {\mathcal B}^+$ holds for commutative $C^*$-algebras
${\mathcal A}$ and $\mathcal B$ which are mathematical frameworks
for classical mechanics. This notion of quantum entanglement has
been one of the key research topics since the nineties, in relation
with possible applications to quantum information theory and quantum computation
theory.

One of the main research topics in the theory of entanglement is to
distinguish entanglement from separability. If we take a rank one
projector onto a product vector $x\otimes y$, then it is easy to see
that its partial transpose is also a rank one projector onto the
product vector $\bar x\otimes y$, where $\bar x$ denotes the vector
whose entries are complex conjugates of the corresponding entries of
the vector $x\in\mathbb C^m$. Recall that the partial transpose
$(X\otimes Y)^\tau$ is given by $X^\ttt\otimes Y$ with the usual
transpose $X^\ttt$ of $X$. Therefore, if $A\in M_m\otimes M_n$ is
separable, then its partial transpose $A^\tau$ is also positive
semi-definite, as was observed by Choi \cite{choi-ppt} and Peres
\cite{peres}. A block matrix $A\in M_m\ot M_n$ is said to be of PPT
(positive partial transpose) if both of $A$ and $A^\tau$ are
positive semi-definite. The notion of PPT turns out be to very
important in quantum physics in relation with bound entanglement.
See Ref. \onlinecite{mpr_horo}.

Woronowicz \cite{woronowicz}  showed that if $m=2$ and $n\le 3$ then
the notions of separability and PPT coincide, and gave an explicit
example of entanglement $A\in M_2\otimes M_4$ which is of PPT. This
kind of block matrix is called a PPTES (positive partial transpose
entangled state) when it is normalized. The first example of PPTES
in $M_3\otimes M_3$ was found by Choi \cite{choi-ppt}. A PPTES $A$
is said to be a PPT entangled edge state, or just an edge state in
short, if there exists no nonzero product vector $x\otimes y\in
{\mathcal R}A$ with $\bar x\ot y\in {\mathcal R}A^\tau$ as was
introduced in Ref. \onlinecite{lkhc}, where ${\mathcal R}A$ denotes
the range space of $A$.  In other words, edge states violate the
range criterion for separability \cite{p-horo} in an extreme way.

Since every PPT state is the convex sum of a separable state and an
edge state, it is essential to classify edge states to understand
the whole structures of PPT states. The first step to classify them
is to use the ranks. A PPT state $A$ is said to be of type $(p,q)$
if the rank of $A$ is $p$ and the rank of $A^\tau$ is $q$, as was
introduced in Ref. \onlinecite{sbl}. Now, we concentrate on the case
of $3\otimes 3$. By the results in Refs.
\onlinecite{chen,hlvc,kye-prod-vec,sko}, we have the following
possibilities of types for $3\otimes 3$ PPT entangled edge states:
\begin{equation}\label{dimension}
(4,4),\ (5,5),\ (6,5),\ (7,5),\ (8,5),\ (6,6),\ (7,6),\ (8,6),
\end{equation}
here we list up types $(p,q)$ with $p\ge q$ by the symmetry. See
Ref. \onlinecite{bdmsst,choi-ppt,clarisse,ha-3,ha-kye-2,ha+kye,p-horo,stormer82}
for concrete examples of $3\otimes 3$ edge states of various types.
We refer to Ref. \onlinecite{kye-prod-vec} for a summary of examples.
All possibilities have been realized  in the literature mentioned
above, except for the case of $(8,6)$. In fact, it has been claimed
in Ref. \onlinecite{sbl} that if there is a $3\otimes 3$ PPT entangled edge
state of type $(p,q)$ then $p+q\le 13$.

The purpose of this note is to present two parameterized examples of
$3\otimes 3$ PPT entangled edge states of type $(8,6)$, to complete
the classification of $3\otimes 3$ edges states by their types.
These examples disprove the above mentioned claim \cite{sbl}.
Our examples also disprove another claim \cite{lein} that if
$D=({\mathcal R}A)^\perp$ and $E=({\mathcal R}A^\tau)^\perp$ for a
PPT state $A\in M_m\ot M_n$ and $\dim D+\dim E=m+n-2$, then there
exist finitely many product vectors $x\ot y \in {\mathcal R}A$ with
$\bar x\ot y\in {\mathcal R}A^\tau$.

After we explain in the next section the notion of PPT edge states
in the context of the whole convex structures of the convex cone generated
by PPT states, we present our construction of two parameterized
examples of edges of type $(8,6)$ in the Section 3. In the last
section, we also exhibit various types of edge states arising
from this construction.

\section{Convex geometry of PPT states}

We denote by $\mathbb V_1$ and $\mathbb T$ the convex cones generated
by all separable and PPT states, respectively. The PPT criterion by Choi and Peres
tells us that the relation $\mathbb V_1\subset\mathbb T$ holds. One
of the best way to understand the whole structures of a given convex
set is to characterize the lattice of all faces.
We have very few general information for the facial structures of the convex cone $\mathbb V_1$ itself.
See Ref. \onlinecite{alfsen} in this direction.
On the other hand, we have an easy way to describe faces of the cone $\mathbb T$
generated by PPT states.

Every faces of the
cone $\mathbb T$ is determined \cite{ha_kye_intersection} by a pair
of subspaces of $\mathbb C^m\otimes\mathbb C^n$. More precisely,
every face of $\mathbb T$ is of the form
$$
\sigma(D,E) =\{A\in\mathbb T: {\mathcal R}A\subseteq D,\ {\mathcal
R}A^\tau\subseteq E\}
$$
for a pair $(D,E)$ of subspaces
of $\mathbb C^m \otimes \mathbb C^n$.
Nevertheless, it is very difficult in general
to determine which pairs of subspaces give rise to faces of the
convex cone $\mathbb T$, and this difficulty is one of the main motivation of this
note. In the case of $2\otimes 2$, all faces of $\mathbb T$ have
been found \cite{ha_kye_intersection} in terms of pairs of subspaces,
using the facial structures \cite{byeon-kye, stormer} of the convex cone of
all positive linear maps between $M_2$.
Recall that a point $x$ of a convex set $C$ is said to be an interior point of $C$
if the line segment from any point of $C$ to $x$ may be extended within $C$. The set of all
interior point of $C$ will be denoted by $\inte C$, which is nothing but the relative interior
of $C$ with respect to the affine manifold generated by $C$. Note that $\inte C$ is never empty
for any convex set $C$. A point of $C$ which is not
an interior point is said to be a boundary point. The set of all boundary points of $C$ will be denoted by $\partial C$.
We recall that the interior of $\sigma(D,E)$ is given by
$$
\inte \sigma(D,E) =\{A\in\mathbb T: {\mathcal R}A= D,\ {\mathcal
R}A^\tau= E\}.
$$

From now on, we compare boundary structures of the two convex cones $\mathbb V_1$ and $\mathbb T$.
Basically, we have the following four cases for a given face $\sigma(D,E)$ of the cone $\mathbb T$:
\begin{itemize}
\item
$\sigma(D,E)\subseteq\mathbb V_1$.
\item
$\sigma(D,E)\nsubseteq\mathbb V_1$ but $\inte\sigma (D,E) \cap \mathbb V_1\neq\emptyset$.
\item
$\inte\sigma (D,E) \cap \mathbb V_1 = \emptyset$ but $\partial \sigma (D,E) \cap \mathbb V_1 \neq \emptyset$.
\item
$\sigma(D,E)\cap \mathbb V_1=\emptyset$.
\end{itemize}
Recall that the range criterion for separability tells us that if a
PPT state $A$ is separable with $D={\mathcal R}A$ and $E={\mathcal
R}A^\tau$ then there exist product vectors $x_\iota\otimes
y_\iota\in\mathbb C^m\otimes\mathbb C^n$ such that
$$
D=\spa\{x_\iota\otimes y_\iota\},\qquad E=\spa\{\bar x_\iota\otimes
y_\iota\}.
$$
We say that a pair $(D,E)$ {\sl satisfies the range criterion} if
there exist product vectors with the above property.
Therefore, we see that if the interior of $\sigma(D,E)$ has a
nonempty intersection with the cone $\mathbb V_1$ then $(D,E)$
satisfies the range criterion. The converse of
this statement is also true as was shown in Ref. \onlinecite{choi_kye}, even
though the converse of the range criterion itself does not hold.
In short, we see that $(D,E)$ satisfies the range criterion if and only if
the first two conditions among the above four hold. In
terms of a PPT state $A$ itself, we see that $({\mathcal R}A,
{\mathcal R}A^\tau)$ satisfies the range criterion if and only if
the smallest face containing $A$ has a separable state in its
interior. Recall that every point $x$ of a convex set determines a unique face
in which $x$ is an interior point. This is the smallest face containing $x$.

It remains two cases to be considered: A face
$\sigma(D,E)$ either touches the cone $\mathbb V_1$ at the boundary
or never touches the cone $\mathbb V_1$. It is easy to see that the
latter case occurs if and only if every element of the face
$\sigma(D,E)$ is a PPT entangled edge state.
If this is the case with $\dim D=p$ and $\dim E=q$ then
every interior point of the face $\sigma(D,E)$ is an edge state of type $(p,q)$, and
every boundary point of $\sigma(D,E)$ is also an edge state of type $(s,t)$
with $s<p$ or $t<q$.

The first step to characterize the lattice of all faces of the cone
$\mathbb T$ is to find all pairs $(p,q)$ of natural numbers for
which there exists a face $\sigma(D,E)$ with $\dim D=p$ and $\dim E=q$.
See Ref. \onlinecite{lein} for this line of research. This classification is
especially important for the cases of separable states and edge
states, since every PPT state is the sum of a separable state and an
edge state. This task for separable states is nothing but to classify the dimensions of pairs of
subspaces satisfying the range criterion.
In the case $2\otimes n$,
all pairs $(p,q)$ of natural numbers have been characterized \cite{choi_kye}
for which there exist pairs $(D,E)$
satisfying the range criterion with $\dim D=p$ and $\dim E=q$.

As for edge states, there are previous results in the literature in
two directions. It was shown  \cite{hlvc,2xn} that if
$A$ is supported on $\mathbb C^m\ot\mathbb C^n$ and
the rank of
$A \in M_m \otimes M_n$ is less than or equal to $\max\{m,n\}$, then
two notions of PPT and separability coincide. This gives a lower
bound for the ranks of $A$ and $A^\tau$ for an edge state $A\in
M_m\otimes M_n$: If $A$ is an $m\ot n$ edge of type $(p,q)$, then we
have
$$
p,q> \max\{m,n\}.
$$
On the other hand, for a given pair $(D,E)$ of subspaces in $\mathbb
C^m\ot\mathbb C^n$, it was shown  \cite{kye-prod-vec} that there
must exist $x\ot y\in D$ with $\bar x\ot y\in E$, whenever either the
inequality
$$
\dim D+\dim E> 2mn-m-n+2
$$
holds, or $\dim D+\dim E= 2mn-m-n+2$ and
$$
\sum_{r+s=m-1}(-1)^r \binom kr\binom \ell s\neq 0
$$
hold with $k=\dim D^\perp$ and $\ell=\dim E^\perp$. This  gives us
an upper bound for the ranks of $A$ and $A^\tau$ for an edge state
$A\in M_m\otimes M_n$.

In case of $m=n=3$, we have $2mn-m-n+2=14$. It is easy to see that
$(k,\ell)=(2,2)$ satisfies the above condition, but $(k,\ell)=(1,3)$
does not satisfy. Furthermore, it is now known \cite{chen,sko}
that every PPT entanglement of rank $4$ is automatically of type
$(4,4)$. All of these arguments give us the possibilities of types
as is given in (\ref{dimension}). See also Ref. \onlinecite{kye-prod-vec} for
the summary in the case of $(m,n)=(2,4)$ as well as in the case of
$m=n=3$. It is unknown whether there exists a $2\otimes 4$ PPT edge
state of type $(6,6)$ or not. Classifications of possible types of
edge states for the $2\otimes 4$ and $3\otimes 3$ cases are summarized in the following pictures:

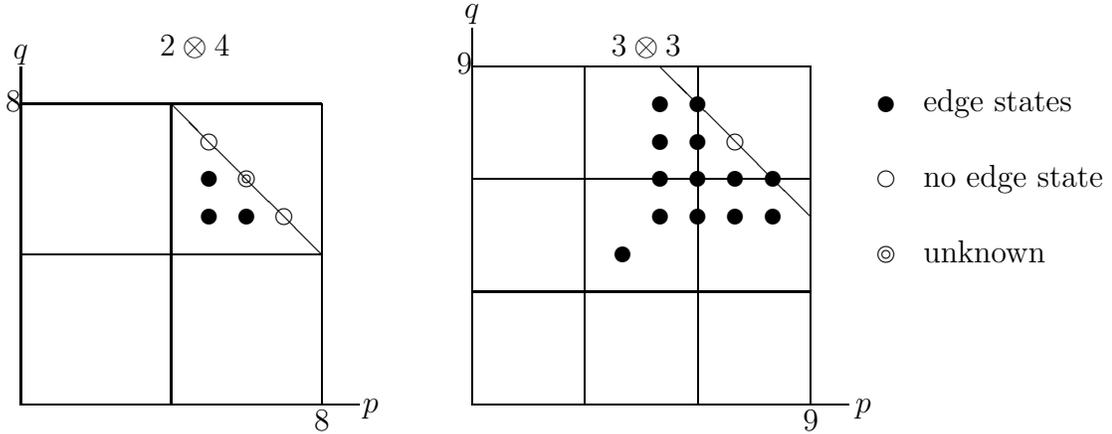
\begin{center}
\setlength{\unitlength}{.5 truecm}
\begin{picture}(26,11)
\put(0,0){\line(1,0){9}} \put(0,0){\line(0,1){9}}
\put(0,8){\line(1,0){8}} \put(8,0){\line(0,1){8}}
\put(0,4){\line(1,0){8}} \put(4,0){\line(0,1){8}}
\put(8,4){\line(-1,1){4}} \put(9.1,-0.2){$p$} \put(-0.2,9.2){$q$}
\put(-0.4,7.8){$8$} \put(7.8,-0.7){$8$}

\newcommand\cii{\circle*{0.4}}

\put(5,5){\cii} \put(5,6){\cii} \put(6,5){\cii}
\put(6,6){\circle{0.2}}\put(6,6){\circle{0.4}}
\put(5,7){\circle{0.4}} \put(7,5){\circle{0.4}}

\put(3.7,9.3){$2\otimes 4$} \put(15.7,9.3){$3\otimes 3$}

\put(12,0){\line(1,0){10}} \put(12,0){\line(0,1){10}}
\put(12,9){\line(1,0){9}} \put(21,0){\line(0,1){9}}
\put(12,3){\line(1,0){9}} \put(15,0){\line(0,1){9}}
\put(12,6){\line(1,0){9}} \put(18,0){\line(0,1){9}}
\put(21,5){\line(-1,1){4}} \put(22.2,-0.2){$p$} \put(11.8,10.3){$q$}
\put(11.6,8.8){$9$} \put(20.8,-0.7){$9$}

\put(16,4){\cii} \put(17,5){\cii} \put(17,6){\cii} \put(17,7){\cii}
\put(17,8){\cii} \put(19,5){\cii} \put(18,5){\cii} \put(20,5){\cii}
\put(18,6){\cii} \put(18,7){\cii} \put(19,6){\cii}
\put(18,8){\cii} \put(20,6){\cii} \put(19,7){\circle{0.4}}

\put(23,8){\circle*{0.4}}\put(24,7.8){edge states}
\put(23,6){\circle{0.4}}\put(24,5.8){no edge state}
\put(23,4){\circle{0.4}}\put(23,4){\circle{0.2}}\put(24,3.8){unknown}
\end{picture}
\end{center}
\medskip

\section{Construction}

We begin with the following $3\times 3$ matrix
$$
P[\theta]:= \left(
\begin{array}{ccccccccccc}
e^{i\theta}+e^{-i\theta}     &-e^{i\theta}     &-e^{-i\theta}    \\
-e^{-i\theta}       &e^{i\theta}+e^{-i\theta}     &-e^{i\theta}     \\
-e^{i\theta}    &-e^{-i\theta}    &e^{i\theta}+e^{-i\theta}
\end{array}
\right)
$$
which has a kernel vector $(1,1,1)^\ttt$. Considering the principal
submatrices, we see that $P[\theta]$ is positive semi-definite if
and only if $\cos\theta\ge 0$ and $2\cos 2\theta\ge -1$ if and only
if $-\frac \pi 3\le\theta\le\frac \pi 3$. If $-\frac \pi
3<\theta<\frac \pi 3$ then $P[\theta]$ is of rank two, and if
$\theta=-\frac \pi 3$ or $\theta=\frac \pi 3$ then $P[\theta]$ is of
rank one.

Consider the following matrix
\begin{equation}\label{def_A}
A= \left(
\begin{array}{ccccccccccc}
e^{i\theta}+e^{-i\theta}     &\cdot   &\cdot       &\cdot  &-e^{i\theta}     &\cdot       &\cdot   &\cdot  &-e^{-i\theta}    \\
\cdot   &\frac 1b &\cdot           &\cdot    &\cdot   &\cdot             &\cdot &\cdot     &\cdot   \\
\cdot  &\cdot    &b         &\cdot &\cdot  &\cdot                 &\cdot    &\cdot &\cdot  \\
\cdot  &\cdot    &\cdot &b &\cdot  &\cdot    &\cdot    &\cdot &\cdot  \\
-e^{-i\theta}     &\cdot   &\cdot  &\cdot  &e^{i\theta}+e^{-i\theta}     &\cdot   &\cdot   &\cdot  &-e^{i\theta}     \\
\cdot   &\cdot &\cdot    &\cdot    &\cdot   &\frac 1b &\cdot &\cdot    &\cdot   \\
\cdot   &\cdot &\cdot    &\cdot    &\cdot   &\cdot &\frac 1b &\cdot    &\cdot   \\
\cdot  &\cdot    &\cdot &\cdot &\cdot  &\cdot    &\cdot    &b &\cdot  \\
-e^{i\theta}     &\cdot   &\cdot  &\cdot  &-e^{-i\theta}     &\cdot
&\cdot   &\cdot  &e^{i\theta}+e^{-i\theta}
\end{array}
\right)
\end{equation}
in $M_3\otimes M_3$ with the conditions
\begin{equation}\label{condition}
b>0,\qquad -\frac \pi 3<\theta<\frac \pi 3,\qquad \theta\neq 0,
\end{equation}
where $\cdot$ denote zero. The partial transpose $A^\tau$ of $A$ is
given by
$$
A^\tau=
\left(
\begin{array}{ccccccccccc}
e^{i\theta}+e^{-i\theta}     &\cdot   &\cdot  &\cdot  &\cdot     &\cdot   &\cdot   &\cdot  &\cdot    \\
\cdot   &\frac 1b &\cdot    &-e^{-i\theta}    &\cdot   &\cdot &\cdot &\cdot     &\cdot   \\
\cdot  &\cdot    &b &\cdot &\cdot  &\cdot    &-e^{i\theta}    &\cdot &\cdot  \\
\cdot  &-e^{i\theta}    &\cdot &b &\cdot  &\cdot    &\cdot    &\cdot &\cdot  \\
\cdot     &\cdot   &\cdot  &\cdot  &e^{i\theta}+e^{-i\theta}      &\cdot   &\cdot   &\cdot  &\cdot     \\
\cdot   &\cdot &\cdot    &\cdot    &\cdot   &\frac 1b &\cdot &-e^{-i\theta}    &\cdot   \\
\cdot   &\cdot &-e^{-i\theta}    &\cdot    &\cdot   &\cdot &\frac 1b &\cdot    &\cdot   \\
\cdot  &\cdot    &\cdot &\cdot &\cdot  &-e^{i\theta}    &\cdot    &b &\cdot  \\
\cdot     &\cdot   &\cdot  &\cdot  &\cdot     &\cdot   &\cdot
&\cdot  &e^{i\theta}+e^{-i\theta}
\end{array}
\right).
$$
It is clear that $A$ is of PPT under the condition
(\ref{condition}), and we have $\rk A=8$ and $\rk A^\tau=6$.

We proceed to show that $A$ is a PPT entangled edge state under the
condition (\ref{condition}). First of all, we note that the kernel
of $A$ is spanned by
$$
(1,0,0\,;\,0,1,0\,;\,0,0,1)^\ttt
$$
and the kernel of $A^\tau$ is spanned by the following three
vectors:
$$
\begin{aligned}
(0,b,0\,;\,e^{i\theta},0,0\,;\,0,0,0)^\ttt,\\
(0,0,0\,;\,0,0,b\,;\,0,e^{i\theta},0)^\ttt,\\
(0,0,e^{i\theta}\,;\,0,0,0\,;\,b,0,0)^\ttt.
\end{aligned}
$$
Suppose that a  product vector $z= x\ot y\in\mathbb C^3\ot\mathbb
C^3$ is in the range of $A$, and $\bar x\ot y$ is in the range of
$A^\tau$. Then we have
\begin{equation}\label{1}
 x_1 y_1+ x_2 y_2+ x_3 y_3=0,
\end{equation}
and
\begin{equation}\label{2}
\begin{aligned}
b \bar x_1 y_2+e^{-i\theta} \bar x_2 y_1=0,\\
b \bar x_2 y_3+e^{-i\theta} \bar x_3 y_2=0,\\
b \bar x_3 y_1+e^{-i\theta} \bar x_1 y_3=0.
\end{aligned}
\end{equation}
From (\ref{2}) we see that at least one of $x_i, y_i$ is zero.
Indeed, we have
$$
b^3 \bar x_1 \bar x_2 \bar x_3 y_1 y_2 y_3=-e^{-3i\theta} \bar x_1
\bar x_2 \bar x_3 y_1 y_2 y_3
$$
by (\ref{2}), from which $\bar x_1 \bar x_2 \bar x_3 y_1 y_2 y_3=0$.
If $x\ot y$ is nonzero, then we also have $ x_i=0
\Longleftrightarrow  y_i=0$ from (\ref{2}).

We first consider the case of $ x_3= y_3=0$. Then we have
$$
 x_1 y_1+ x_2 y_2=0,\qquad b \bar x_1 y_2+e^{-i\theta} \bar x_2 y_1=0,
$$
from which we have
$$
 \bar x_1 x_1 y_1+ \bar x_1 x_2 y_2=0,\qquad b \bar x_1 x_2
y_2+e^{-i\theta} \bar x_2 x_2 y_1=0.
$$
Therefore, we get
$$
| x_1|^2 y_1=- \bar x_1 x_2 y_2=\frac{e^{-i\theta}}{b}| x_2|^2 y_1.
$$
Since $\theta\neq 0$, we conclude that $ x_1= x_2=0$ or $ y_1=0$. If
$ x_1= x_2=0$, then $ x=0$. If $ y_1=0$ and either $ x_1$ or $ x_2$
is nonzero, then we have $ y=0$. Similar arguments for the cases
$x_1=y_1=0$ and $x_2=y_2=0$ show that if $ x, y\in\mathbb C^3$
satisfy the relations (\ref{1}) and (\ref{2}), then $ x\ot y=0$.
This shows that there exists no nonzero product vector $ x\ot y\in
{\mathcal R}A$ with $ \bar x\ot y\in{\mathcal R}A^\tau$. Therefore,
we conclude that $A$ is a PPT entangled edge state of type $(8,6)$.

Recall \cite{wall} that every $5$-dimensional subspace of $\mathbb C^3\otimes\mathbb C^3$ has a product vector.
This is equivalent to say that every system of equations consisting of four homogeneous linear
equations with respect to unknowns $\{x_iy_j: i,j=1,2,3\}$
must have nontrivial solutions. But, the system of four equations from (\ref{1}) and (\ref{2}) involve complex conjugates,
and may not have nonzero solutions. This seems to be the main point for the wrong statements in
Ref. \onlinecite{sbl,lein}.

For nonnegative real numbers $a,b$ and $c$, we consider the following linear map
$$
\Phi[a,b,c](X)=\\
\begin{pmatrix}
ax_{11}+bx_{22}+cx_{33} & -x_{12} & -x_{13} \\
-x_{21} & cx_{11}+ax_{22}+bx_{33} & -x_{23} \\
-x_{31} & -x_{32} & bx_{11}+cx_{22}+ax_{33}
\end{pmatrix}
$$
 between $M_3$, as was introduced in Ref. \onlinecite{cho-kye-lee}. We also recall that
the Choi matrix $C_\phi\in M_m\otimes M_n$ of a linear map $\phi:M_m\to M_n$ is given by
$$
C_\phi:=\sum_{i,j=1}^m e_{ij}\otimes\phi(e_{ij})\in M_m\otimes M_n,
$$
and $C_\phi$ is of PPT if and only if $\phi$ is both completely
positive and completely copositive by Ref. \onlinecite{choi75-10}.
We also note that $\Phi[a,b,c]$ is both completely positive and completely copositive
if and only if $a\ge 2$ and $bc\ge 1$ by Ref. \onlinecite{cho-kye-lee}. If $\theta
=0$ then the matrix $A$ in (\ref{def_A}) is just the Choi matrix of
the map $\Phi[2,b,\frac 1b]$, which is a PPT state of type $(8,6)$.
On the other hand, we have the following PPT states
\begin{equation}\label{7_6}
A= \left(
\begin{array}{ccccccccccc}
1     &\cdot   &\cdot       &\cdot  &1     &\cdot       &\cdot   &\cdot  &1   \\
\cdot   &\frac 1b &\cdot           &\cdot    &\cdot   &\cdot             &\cdot &\cdot     &\cdot   \\
\cdot  &\cdot    &b         &\cdot &\cdot  &\cdot                 &\cdot    &\cdot &\cdot  \\
\cdot  &\cdot    &\cdot &b &\cdot  &\cdot    &\cdot    &\cdot &\cdot  \\
1     &\cdot   &\cdot  &\cdot  &1     &\cdot   &\cdot   &\cdot  &1     \\
\cdot   &\cdot &\cdot    &\cdot    &\cdot   &\frac 1b &\cdot &\cdot    &\cdot   \\
\cdot   &\cdot &\cdot    &\cdot    &\cdot   &\cdot &\frac 1b &\cdot    &\cdot   \\
\cdot  &\cdot    &\cdot &\cdot &\cdot  &\cdot    &\cdot    &b &\cdot  \\
1     &\cdot   &\cdot  &\cdot  &1     &\cdot    &\cdot   &\cdot  &1
\end{array}
\right)
\end{equation}
of type $(7,6)$ in the literature \cite{ha-kye-2},
which is an edge state
whenever $b>0$ and $b\neq 1$.
The key idea of the construction was to parameterized
offdiagonals $-1$ and $1$ of these two cases by $e^{i\theta}$.
We note that a variant of (\ref{7_6}) has been used by
St\o rmer \cite{stormer82} to give a short proof of the indecomposability of the Choi
map $\Phi[1,0,\lambda]$ for $\lambda \ge 1$.

If $\theta=0$ then $A$
in (\ref{def_A})
turns out to be separable. Indeed, if we take product vectors
$$
\begin{aligned}
z_1(\omega)&=(0,1,\sqrt b\,\omega)\otimes (0,\sqrt b,-\bar\omega)
=(0,0,0 \,;\, 0,\sqrt b, -\bar\omega \,;\, 0,b\, \omega,-\sqrt b\,)\\
z_2(\omega)&=(\sqrt b\,\omega,0,1)\otimes (-\bar\omega,0,\sqrt b)
=(-\sqrt b,0,b\, \omega \,;\, 0,0,0 \,;\, -\bar\omega,0,\sqrt b\,)\\
z_3(\omega)&=(1,\sqrt b\,\omega,0)\otimes (\sqrt b,-\bar\omega,0)
=(\sqrt b,-\bar\omega,0 \,;\, b\, \omega,-\sqrt b,0 \,;\, 0,0,0)
\end{aligned}
$$
in $\mathbb C^3\otimes\mathbb C^3$ then it is straightforward to see that
$$
A=\dfrac 1{3b}
\sum_{i=1}^3\sum_{\omega\in\Omega}z_i(\omega)z_i(\omega)^*,
$$
where
$\Omega=\{1,e^{\frac 23\pi i}, e^{-\frac 23\pi i}\}$ is the third roots of unity.
We note that the Choi matrix of the map $\Phi[a,b,c]$ is of PPT if and only if $a\ge 2$ and $bc\ge 1$, and so it is the sum of
a diagonal matrix with nonnegative diagonal entries and a separable one.
Therefore, we see that the Choi matrix of the map $\Phi[a,b,c]$ is of PPT if and only if it is separable.
This shows that the linear map $\Phi[a,b,c]$ is super-positive in the sense of Ref. \onlinecite{ando},
or equivalently an entanglement breaking channel in the sense of Ref. \onlinecite{hsrus,hol}
if and only if it is both completely positive and completely copositive if and only if
$a\ge 2$ and $bc\ge 1$.
See Ref. \onlinecite{singh} for related topics.

If we put the following number
$$
a_\theta=\max\{
e^{i(\theta+\frac 32\pi)}+e^{-i(\theta+\frac 32\pi)},\
e^{i\theta}+e^{-i\theta},\
e^{i(\theta-\frac 32\pi)}+e^{-i(\theta-\frac 32\pi)}\}
$$
in the place of $e^{i\theta}+e^{-i\theta}$ when we define the matrix $A$
in (\ref{def_A}),
then we have
similar PPT edge states for every $\theta$. Note that $a_\theta$ is the smallest number so that
$$
\left(
\begin{array}{ccccccccccc}
a_\theta     &-e^{i\theta}     &-e^{-i\theta}    \\
-e^{-i\theta}       &a_\theta    &-e^{i\theta}     \\
-e^{i\theta}    &-e^{-i\theta}    &a_\theta
\end{array}
\right)
$$
is positive semi-definite.

\section{Edge states of other types}

Let $A$ be the matrix given by (\ref{def_A}).
Now, we search edge states $X$ in the smallest face containing $A$
by a similar method as in Ref. \onlinecite{ha-kye-2}.
Note that $X$ is in this face if and only if the relations
$$
{\mathcal R}X\subseteq {\mathcal R}A,\qquad {\mathcal R}X^\tau\subseteq
{\mathcal R}A^\tau
$$
hold. Note that every range vector of $A^\tau$ is of the form
$$
\left(\xi_i,\frac{\alpha_i}{\sqrt b}, -\gamma_i \sqrt be^{i\theta}\,;\,
-\alpha_i \sqrt b e^{i\theta},\eta_i,\frac{\beta_i}{\sqrt b}\,;\,
\frac{\gamma_i}{\sqrt b},-\beta_i\sqrt b e^{i\theta},\zeta_i\right)^\ttt,
$$
for scalars $\xi_i,\eta_i,\zeta_i,\alpha_i,\beta_i$ and $\gamma_i$.
We denote by $P$ the rank one projector onto the vector
$$
\left(1,\frac{1}{\sqrt b}, - \sqrt be^{i\theta}\,;\,
- \sqrt b e^{i\theta},1,\frac{1}{\sqrt b}\,;\,
\frac{1}{\sqrt b},-\sqrt b e^{i\theta},1\right)^\ttt,
$$
and by $Q_i$ onto the vector
$$
\left(\xi_i,\alpha_i, \gamma_i \,;\,
\alpha_i ,\eta_i,\beta_i\,;\,
\gamma_i,\beta_i,\zeta_i\right)^\ttt,
$$
respectively. Here, the projector onto a column vector $w$ means the rank one matrix $ww^*$.
Then we see that $A^\tau$ is the Hadamard product of $P$ and $\sum_i Q_i$ for suitable
choice of $\xi_i,\eta_i,\zeta_i,\alpha_i,\beta_i$ and $\gamma_i$.
If we write $\xi,\eta,\zeta,\alpha,\beta$ and $\gamma$ the vectors
whose $i$-th entries are $\xi_i,\eta_i,\zeta_i,\alpha_i,\beta_i$ and
$\gamma_i$, respectively, then
the matrix $X=(X^\tau)^\tau$ is
the Hadamard product of the following two matrices:
$$
\left(
\begin{array}{ccccccccccc}
1 &\frac 1{\sqrt b}  &-{\sqrt b}e^{-i\theta}&
&-\sqrt be^{i\theta} &-e^{i\theta}  &b &
&\frac 1{\sqrt b}  &\frac 1b   &-e^{-i\theta}
\\
\frac 1{\sqrt b}  &\frac 1b   &-e^{-i\theta}  &
&1 &\frac 1{\sqrt b}  &-{\sqrt b}e^{-i\theta}&
&-\sqrt be^{i\theta} &-e^{i\theta}  &b
\\
-\sqrt be^{i\theta} &-e^{i\theta}  &b &
&\frac 1{\sqrt b}  &\frac 1b   &-e^{-i\theta}  &
&1 &\frac 1{\sqrt b}  &-{\sqrt b}e^{-i\theta}
\\
\\
-{\sqrt b}e^{-i\theta}  &1 &\frac 1{\sqrt b} &
 &b &-\sqrt be^{i\theta} &-e^{i\theta} &
&-e^{-i\theta}   &\frac 1{\sqrt b}  &\frac 1b
\\
-e^{-i\theta}   &\frac 1{\sqrt b}  &\frac 1b &
&-{\sqrt b}e^{-i\theta}  &1 &\frac 1{\sqrt b} &
 &b &-\sqrt be^{i\theta} &-e^{i\theta}
\\
 b &-\sqrt be^{i\theta} &-e^{i\theta} &
&-e^{-i\theta}   &\frac 1{\sqrt b}  &\frac 1b &
&-{\sqrt b}e^{-i\theta}  &1 &\frac 1{\sqrt b}
\\
\\
\frac 1{\sqrt b} &-{\sqrt b}e^{-i\theta} &1&
 &-e^{i\theta}  &b &-\sqrt be^{i\theta}&
&\frac 1b  &-e^{-i\theta}  &\frac 1{\sqrt b}
\\
\frac 1b  &-e^{-i\theta}  &\frac 1{\sqrt b}&
&\frac 1{\sqrt b} &-{\sqrt b}e^{-i\theta} &1&
 &-e^{i\theta}  &b &-\sqrt be^{i\theta}
\\
 -e^{i\theta}  &b &-\sqrt be^{i\theta}&
 &\frac 1b  &-e^{-i\theta}  &\frac 1{\sqrt b}&
&\frac 1{\sqrt b} &-{\sqrt b}e^{-i\theta} &1
\\
\end{array}
\right)
$$
and
$$
\left(
\begin{array}{ccccccccccc}
(\xi|\xi)      &(\xi|\alpha)      &(\xi|\gamma)       &&(\alpha|\xi)   &(\alpha|\alpha)   &(\alpha|\gamma)  &&(\gamma|\xi)   &(\gamma|\alpha)   &(\gamma|\gamma)\\
(\alpha|\xi)   &(\alpha|\alpha)   &(\alpha|\gamma)  &&(\eta|\xi)     &(\eta|\alpha)     &(\eta|\gamma)    &&(\beta|\xi)    &(\beta|\alpha)    &(\beta|\gamma)   \\
(\gamma|\xi)   &(\gamma|\alpha)   &(\gamma|\gamma)  &&(\beta|\xi)    &(\beta|\alpha)    &(\beta|\gamma)   &&(\zeta|\xi)    &(\zeta|\alpha)      &(\zeta|\gamma) \\
\\
(\xi|\alpha)  &(\xi|\eta) &(\xi|\beta)               &&(\alpha|\alpha)   &(\alpha|\eta)  &(\alpha|\beta)    &&(\gamma|\alpha)  &(\gamma|\eta) &(\gamma|\beta)  \\
(\alpha|\alpha)   &(\alpha|\eta)  &(\alpha|\beta)     &&(\eta|\alpha)  &(\eta|\eta) &(\eta|\beta)            &&(\beta|\alpha)  &(\beta|\eta) &(\beta|\beta) \\
(\gamma|\alpha)  &(\gamma|\eta) &(\gamma|\beta)       &&(\beta|\alpha)  &(\beta|\eta) &(\beta|\beta)          &&(\zeta|\alpha)  &(\zeta|\eta) &(\zeta|\beta) \\
\\
(\xi|\gamma) &(\xi|\beta) &(\xi|\zeta)               &&(\alpha|\gamma)  &(\alpha|\beta)  &(\alpha|\zeta)   &&(\gamma|\gamma) &(\gamma|\beta) &(\gamma|\zeta) \\
(\alpha|\gamma)  &(\alpha|\beta)  &(\alpha|\zeta)   &&(\eta|\gamma) &(\eta|\beta) &(\eta|\zeta)         &&(\beta|\gamma) &(\beta|\beta) &(\beta|\zeta) \\
(\gamma|\gamma) &(\gamma|\beta) &(\gamma|\zeta)    &&(\beta|\gamma) &(\beta|\beta) &(\beta|\zeta)           &&(\zeta|\gamma) &(\zeta|\beta) &(\zeta|\zeta) \\
\end{array}
\right).
$$

Since $(1,0,0\,;\,0,1,0\,;\,0,0,1)^\ttt$ is in the $\ker A\subseteq\ker X$, we have
$$
\|\xi\|^2=e^{i\theta}\|\alpha\|^2+e^{-i\theta}\|\gamma\|^2,\quad
\|\eta\|^2=e^{i\theta}\|\beta\|^2+e^{-i\theta}\|\alpha\|^2,\quad
\|\zeta\|^2=e^{i\theta}\|\gamma\|^2+e^{-i\theta}\|\beta\|^2,
$$
and so we have $\|\alpha\|=\|\beta\|=\|\gamma||$.
If $\|\alpha\|=\|\beta\|=\|\gamma||=0$, then $A=0$. So, we may assume that
$$\|\alpha\|=\|\beta\|=\|\gamma||=1.
$$
Then we have
\begin{equation}\label{con1}
\|\xi\|^2=\|\eta\|^2=\|\zeta\|^2=e^{i\theta}+e^{-i\theta}.
\end{equation}
Considering $(2,4)$, $(6,8)$ and $(7,3)$ principal submatrices, we also have
\begin{equation}\label{con2}
|(\xi|\eta)|\le 1, \qquad
|(\eta|\zeta)|\le 1, \qquad
|(\zeta|\xi)|\le 1.
\end{equation}

If we take vectors so that
$\spa\{\xi,\eta,\zeta\}\perp\spa\{\alpha,\beta,\gamma\}$ with
mutually orthonormal vectors $\alpha,\beta,\gamma$ then we have
$$
X= \left(
\begin{array}{ccccccccccc}
e^{i\theta}+e^{-i\theta}      &\cdot      &\cdot       &&\cdot   &-e^{i\theta}   &\cdot  &&\cdot   &\cdot   &-e^{-i\theta}\\
\cdot   &\frac 1b   &\cdot  &&(\eta|\xi)     &\cdot     &\cdot    &&\cdot    &\cdot    &\cdot   \\
\cdot   &\cdot   &b  &&\cdot    &\cdot    &\cdot   &&(\zeta|\xi)    &\cdot      &\cdot \\
\\
\cdot  &(\xi|\eta) &\cdot               &&b   &\cdot  &\cdot    &&\cdot  &\cdot &\cdot  \\
-e^{-i\theta}   &\cdot  &\cdot     &&\cdot  &e^{i\theta}+e^{-i\theta} &\cdot            &&\cdot  &\cdot &-e^{i\theta} \\
\cdot  &\cdot &\cdot      &&\cdot  &\cdot &\frac 1b          &&\cdot  &(\zeta|\eta) &\cdot \\
\\
\cdot &\cdot &(\xi|\zeta)               &&\cdot  &\cdot  &\cdot   &&\frac 1b &\cdot &\cdot \\
\cdot  &\cdot  &\cdot  &&\cdot &\cdot &(\eta|\zeta)         &&\cdot &b&\cdot \\
-e^{i\theta} &\cdot &\cdot    &&\cdot &-e^{-i\theta} &\cdot           &&\cdot &\cdot &e^{i\theta}+e^{-i\theta} \\
\end{array}
\right)
$$
and
$$
X^\tau=
\left(
\begin{array}{ccccccccccc}
e^{i\theta}+e^{-i\theta}      &\cdot      &\cdot       &&\cdot   &(\xi|\eta)   &\cdot  &&\cdot   &\cdot   &(\xi|\zeta)\\
\cdot   &\frac 1b   &\cdot  &&-e^{-i\theta}     &\cdot     &\cdot    &&\cdot    &\cdot    &\cdot   \\
\cdot   &\cdot   &b  &&\cdot    &\cdot    &\cdot   &&-e^{i\theta}    &\cdot      &\cdot \\
\\
\cdot  &-e^{i\theta} &\cdot               &&b   &\cdot  &\cdot    &&\cdot  &\cdot &\cdot  \\
(\eta|\xi)   &\cdot  &\cdot     &&\cdot  &e^{i\theta}+e^{-i\theta} &\cdot            &&\cdot  &\cdot &(\eta|\zeta) \\
\cdot  &\cdot &\cdot      &&\cdot  &\cdot &\frac 1b          &&\cdot  &-e^{-i\theta} &\cdot \\
\\
\cdot &\cdot &-e^{-i\theta}               &&\cdot  &\cdot  &\cdot   &&\frac 1b &\cdot &\cdot \\
\cdot  &\cdot  &\cdot  &&\cdot &\cdot &-e^{i\theta}         &&\cdot &b&\cdot \\
(\zeta|\xi) &\cdot &\cdot    &&\cdot &(\zeta|\eta) &\cdot           &&\cdot &\cdot &e^{i\theta}+e^{-i\theta} \\
\end{array}
\right).
$$
It is clear that $X$ is of PPT under the conditions (\ref{con1}) and (\ref{con2}).
We note that the rank of $X$ is equal to
$$
2+\rk\left(\begin{matrix}\frac 1b&(\xi|\eta)\\(\eta|\xi)&b\end{matrix}\right)
+\rk\left(\begin{matrix}\frac 1b&(\eta|\zeta)\\(\zeta|\eta)&b\end{matrix}\right)
+\rk\left(\begin{matrix}b&(\zeta|\xi)\\(\xi|\zeta)&\frac 1b\end{matrix}\right)
$$
and the rank of $X^\tau$ is equal to
$$
3+\rk\left(\begin{matrix}
e^{i\theta}+e^{-i\theta} & (\xi|\eta) & (\xi|\zeta)\\
(\eta|\xi) &e^{i\theta}+e^{-i\theta}& (\eta|\zeta)\\
(\zeta|\xi) &(\zeta|\eta)&e^{i\theta}+e^{-i\theta}\end{matrix}\right)
=
3+\dim\spa\{\xi,\eta,\zeta\}.
$$

In the three dimensional space $\mathbb C^3$, it is possible to take
linearly independent vectors $\xi,\eta,\zeta$ satisfying (\ref{con1}) and (\ref{con2}) so that
some of
$$
(\xi|\eta),\qquad (\eta|\zeta),\qquad (\zeta|\xi)
$$
are of absolute values one and the remainders are zero. Therefore, we get examples
of edge states of types
$(8,6), (7,6), (6,6)$ and $(5,6)$.

To get edge states of type $(p,5)$ for $p=5,6,7,8$, it is convenient to consider the matrix
$$
P[\rho,\sigma,\tau]:=\left(
\begin{array}{ccccccccccc}
e^{i\theta}+e^{-i\theta}     &\rho     &\bar \tau    \\
\bar \rho       &e^{i\theta}+e^{-i\theta}     &\sigma     \\
\tau    &\bar \sigma    &e^{i\theta}+e^{-i\theta}
\end{array}
\right)
$$
with the conditions
\begin{equation}\label{ugkh}
|\rho|\le 1,\qquad |\sigma|\le 1,\qquad |\tau|\le 1,\qquad \det P[\rho,\sigma,\tau]=0.
\end{equation}
Then $P[\rho,\sigma,\tau]$ is a positive semi-definite matrix of rank two. By
spectral decomposition, we may get two vectors
$E_1=(\xi_1,\eta_1,\zeta_1)$ and $E_2=(\xi_2,\eta_2,\zeta_2)$ so
that $P[\rho,\sigma,\tau]$ is the sum of rank one projectors onto $E_1$ and
$E_2$, respectively. Then we see that
$$
P[\rho,\sigma,\tau]=\left(
\begin{array}{ccccccccccc}
|\xi_1|^2     &\xi_1\bar\eta_1    &\xi_1\bar\zeta_1   \\
\eta_1\bar\xi_1     &|\eta_1|^2     &\eta_1\bar\zeta_1   \\
\zeta_1\bar\xi_1 &\zeta_1\bar\eta_1   &|\zeta_1|^2
\end{array}
\right)
+
\left(
\begin{array}{ccccccccccc}
|\xi_2|^2     &\xi_2\bar\eta_2    &\xi_2\bar\zeta_2   \\
\eta_2\bar\xi_2     &|\eta_2|^2     &\eta_2\bar\zeta_2   \\
\zeta_2\bar\xi_2 &\zeta_2\bar\eta_2   &|\zeta_2|^2
\end{array}
\right)
=
\left(\begin{matrix}
(\xi|\xi) & (\xi|\eta) & (\xi|\zeta)\\
(\eta|\xi) &(\eta|\eta)& (\eta|\zeta)\\
(\zeta|\xi) &(\zeta|\eta)&(\zeta|\zeta)\end{matrix}\right).
$$
If we take $\rho,\sigma,\tau$ with (\ref{ugkh}) so that some of them are of absolute values one and the remainders of them have
the absolute values less than one, then we get PPT entangled states of types
$(8,5), (7,5), (6,5)$ and $(5,5)$, as we will now show.

For a given fixed $\theta$ with (\ref{condition}), we can take a real number $r$ with $-1<r<1$ so that
$$
P[r,r,r],\qquad P[r,-r,1],\qquad P[1,1,r]
$$
is of rank two, respectively, to get edge states of types $(8,5), (7,5)$ and $(6,5)$.
For example, we see that
$$
P[-\cos\theta,-\cos\theta,-\cos\theta]
$$
is of rank two, and so
we get the following natural examples of $3\otimes 3$ edge states of type $(8,5)$:
$$
\left(
\begin{array}{ccccccccccc}
e^{i\theta}+e^{-i\theta}      &\cdot      &\cdot       &&\cdot   &-e^{i\theta}   &\cdot  &&\cdot   &\cdot   &-e^{-i\theta}\\
\cdot   &\frac 1b   &\cdot  &&-\cos\theta     &\cdot     &\cdot    &&\cdot    &\cdot    &\cdot   \\
\cdot   &\cdot   &b  &&\cdot    &\cdot    &\cdot   &&-\cos\theta    &\cdot      &\cdot \\
\\
\cdot  &-\cos\theta &\cdot               &&b   &\cdot  &\cdot    &&\cdot  &\cdot &\cdot  \\
-e^{-i\theta}   &\cdot  &\cdot     &&\cdot  &e^{i\theta}+e^{-i\theta} &\cdot            &&\cdot  &\cdot &-e^{i\theta} \\
\cdot  &\cdot &\cdot      &&\cdot  &\cdot &\frac 1b          &&\cdot  &-\cos\theta &\cdot \\
\\
\cdot &\cdot &-\cos\theta               &&\cdot  &\cdot  &\cdot   &&\frac 1b &\cdot &\cdot \\
\cdot  &\cdot  &\cdot  &&\cdot &\cdot &-\cos\theta         &&\cdot &b&\cdot \\
-e^{i\theta} &\cdot &\cdot    &&\cdot &-e^{-i\theta} &\cdot           &&\cdot &\cdot &e^{i\theta}+e^{-i\theta} \\
\end{array}
\right).
$$
To get examples of edge states of types $(7,5)$ and $(6,5)$, we put $\omega=e^{i\theta}+e^{-i\theta}$ temporarily. Note that $1<\omega<2$.
We also note that
$$
\det P[r,-r,1]=(1+\omega)(\omega^2-\omega-2r^2),\qquad
\det P[1,1,r]=(\omega-r)(r\omega+\omega^2-2).
$$
and zeros
$$
r=\sqrt{\dfrac{\omega^2-\omega}2}=\sqrt{2\cos^2\theta-\cos\theta},\qquad
r=\dfrac {2-\omega^2}\omega=-\dfrac{\cos 2\theta}{\cos\theta}
$$
of them are in the interval $(-1,1)$, respectively. In this way, we get edge states of types $(7,5)$ and $(6,5)$.
If we consider the rank two matrix $P[-e^{i\theta},-e^{i\theta},-e^{i\theta}]$, which is nothing but $P[\theta]$
at the beginning of the construction,
then we have the following parameterized example of edge states of type $(5,5)$:
$$
\left(
\begin{array}{ccccccccccc}
e^{i\theta}+e^{-i\theta}     &\cdot   &\cdot  &\cdot  &-e^{i\theta}     &\cdot   &\cdot   &\cdot  &-e^{-i\theta}    \\
\cdot   &\frac 1b &\cdot    &-e^{-i\theta}    &\cdot   &\cdot &\cdot &\cdot     &\cdot   \\
\cdot  &\cdot    &b &\cdot &\cdot  &\cdot    &-e^{i\theta}    &\cdot &\cdot  \\
\cdot  &-e^{i\theta}    &\cdot &b &\cdot  &\cdot    &\cdot    &\cdot &\cdot  \\
-e^{-i\theta}     &\cdot   &\cdot  &\cdot  &e^{i\theta}+e^{-i\theta}     &\cdot   &\cdot   &\cdot  &-e^{i\theta}     \\
\cdot   &\cdot &\cdot    &\cdot    &\cdot   &\frac 1b &\cdot &-e^{-i\theta}    &\cdot   \\
\cdot   &\cdot &-e^{-i\theta}    &\cdot    &\cdot   &\cdot &\frac 1b &\cdot    &\cdot   \\
\cdot  &\cdot    &\cdot &\cdot &\cdot  &-e^{i\theta}    &\cdot    &b &\cdot  \\
-e^{i\theta}     &\cdot   &\cdot  &\cdot  &-e^{-i\theta}     &\cdot   &\cdot   &\cdot  &e^{i\theta}+e^{-i\theta}
\end{array}
\right).
$$

In conclusion, we have constructed $3\otimes 3$ PPT entangled edge
states of type $(8,6)$ whose existence has been a long-standing
question since the claim in  Ref. \onlinecite{sbl} without proof. In
this vein, it would be also an interesting question whether there
exists a $2\otimes 4$ edge states of type $(6,6)$ or not, as was
explained in Ref. \onlinecite{kye-prod-vec}. We have shown that
there exist edge states of all possible types in the face generated
by each PPT state we constructed, except for edge states of $(4,4)$
types. These include parameterized examples of edge states of types
$(5,5)$ and $(6,6)$, for which there have been known very few
discrete examples \cite{clarisse,ha-3}. We also have natural
parameterized examples of edge states of type $(8,5)$. Compare with
Ref. \onlinecite{ha-kye-2}. We note that the study of bi-qutrit edge
states with minimal ranks was initiated by  Ref.
\onlinecite{bdmsst}, and have been recently studied in  Ref.
\onlinecite{chen,sko,hhms,slm} very extensively. It is the authors'
hope that this is the starting point for the further study of
bi-qutrit edge states with maximal ranks.

SHK was partially supported by NRFK 2011-0001250. HO was  partially
supported by the JSPS grant for Scientific Research No.20540220. The
first author is grateful to Kil-Chan Ha for helpful discussion.


\begin{thebibliography}{999}


\bibitem{alfsen}
E. Alfsen and F. Shultz,
\it Unique decompositions, faces, and automorphisms of separable states,
\rm J. Math. Phys. \bf 51\rm (2010), 052201.

\bibitem{ando}
T. Ando,
\it Cones and norms in the tensor product of matrix spaces,
\rm  Linear Algebra Appl. \bf 379 \rm (2004), 3--41.

\bibitem{bdmsst}
C. H. Bennett, D. P. DiVincenzo, T. Mor, P. W. Shor, J. A. Smolin,
and B. M. Terhal, \it Unextendible product bases and bound
entanglement,
\rm Phys. Rev. Lett. \bf 82 \rm (1999), 5385--5388.

\bibitem{byeon-kye}
E.-S. Byeon and S.-H. Kye,
\it Facial structures for positive linear maps in the two dimensional matrix algebra,
\rm Positivity \bf 6 \rm (2002), 369--380.

\bibitem{chen}
L. Chen and D. \v Z. Djokovi\'c, \it Description of rank four PPT
entangled states of two qutrits, \rm J. Math. Phys. \bf 52 \rm (2011), 122203.

\bibitem{cho-kye-lee}
S.-J. Cho, S.-H. Kye, and S. G. Lee, \it Generalized Choi maps in
$3$-dimensional matrix algebras, \rm Linear Alg. Appl. \bf 171 \rm (1992), 213--224.

\bibitem{choi_kye}
H.-S. Choi and S.-H. Kye, \it Facial structures for separable
states, \rm J. Korean Math. Soc., \bf 49 \rm (2012), to appear,
\texttt{http://www.math.snu.ac.kr/{$\sim$}kye/paper/separable.pdf}.

\bibitem{choi75-10}
M.-D. Choi, \it Completely positive linear maps on complex matrices,
\rm Linear Alg. Appl. \bf 10 \rm  (1975), 285--290.

\bibitem{choi-ppt}
M.-D. Choi,
\it Positive linear maps,
\rm Operator Algebras and Applications (Kingston, 1980), pp. 583--590,
Proc. Sympos. Pure Math. Vol 38. Part 2, Amer. Math. Soc., 1982.

\bibitem{clarisse}
L. Clarisse, \it Construction of bound entangled edge states with
special ranks,
\rm Phys. Lett. A \bf 359 \rm (2006), 603--607.

\bibitem{ha-3}
K.-C. Ha, \it Comment on : \lq\lq Construction of bound entangled
edge states with special ranks\rq\rq\ \rm[Phys. Lett. A 359 (2006)
603],
\rm Phys. Lett. A \bf 361 \rm (2007), 515--519.

\bibitem{ha_kye_intersection}
K.-C. Ha and S.-H. Kye, \it Construction of entangled states with
positive partial transposes based on indecomposable positive linear
maps,
\rm Phys. Lett. A \bf 325 \rm (2004), 315--323.

\bibitem{ha-kye-2}
K.-C. Ha and S.-H. Kye, \it Construction of $3 \otimes 3$ entangled
edge states with positive partial transposes,
\rm J. Phys. A \bf 38 \rm (2005), 9039--9050.

\bibitem{ha+kye}
K.-C. Ha, S.-H. Kye, and Y. S. Park, \it Entanglements with positive
partial transposes arising from indecomposable positive linear maps,
\rm Phys. Lett. A \bf 313 \rm (2003), 163--174.

\bibitem{hhms}
L. O. Hansen, A. Hauge, J. Myrheim, and P. \O . Sollid, \it Low rank
positive partial transpose states and their relation to product
vectors, \rm preprint, arXiv:1104.1519.

\bibitem{mpr_horo}
M. Horodecki, P. Horodecki and R. Horodecki, \it Mixed-state
entanglement and distillation: is there a ``bound'' entanglement in
nature?, \rm Phys. Rev. Lett. \bf 80 \rm (1998), 5239--5242.

\bibitem{hsrus}
M. Horodecki, P. W. Shor and M. B. Ruskai,
\it General entanglement braking channels,
\rm Rev. Math. Phys. \bf 15 \rm (2003), 629--641.

\bibitem{p-horo}
P. Horodecki, \it Separability criterion and inseparable mixed
states with positive partial transposition,
\rm Phys. Lett. A \bf 232 \rm (1997), 333--339.

\bibitem{hlvc}
P. Horodecki, M. Lewenstein, G. Vidal, and I. Cirac, \it Operational
criterion and constructive checks for the separability of low rank
density matrices,
\rm Phys. Rev. A \bf 62 \rm (2000), 032310.

\bibitem{kye-prod-vec}
Y.-H. Kiem, S.-H. Kye, and J. Lee, \it Existence of product vectors
and their partial conjugates in a pair of spaces,
\rm J. Math. Phys, \bf 52 \rm (2011), 122201.

\bibitem{hol}
A. S. Kholevo, M. E. Shirokov and R. F. Werner,
\it On the notion of entanglement in Hilbert spaces,
\rm Russian Math. Surveys \bf 60 \rm (2005), 359--360.

\bibitem{lein}
J. M. Leinaas, J. Myrheim, and P. \O . Sollid, \it Numerical studies
of entangled PPT states in composite quantum systems,
\rm Phys. Rev. A \bf 81 \rm (2010), 062329.

\bibitem{lkhc}
M. Lewenstein, B. Kraus, P. Horodecki, and J. I. Cirac, \it
Characterization of separable states and entanglement witnesses,
\rm Phys. Rev. A \bf 63 \rm (2001), 044304.

\bibitem{2xn}
B. Kraus, J. I. Cirac, S. Karnas, and M. Lewenstein, \it
Separability in \rm 2xN \it composite quantum systems,
\rm Phys. Rev. A \bf 61 \rm (2000), 062302.

\bibitem{peres}
A. Peres, \it Separability criterion for density matrices,
\rm Phys. Rev. Lett. \bf 77 \rm (1996), 1413--1415.

\bibitem{sbl}
A. Sanpera, D. Bru\ss, and M. Lewenstein,
\it Schmidt-number witnesses and bound entanglement,
\rm Phys. Rev. A \bf 63 \rm (2001), 050301.

\bibitem{singh}
A. I. Singh,
\it Quantum Dynamical Semigroups involving Separable and Entangled States,
\rm preprint,
arXiv:1201.0250.

\bibitem{sko}
\L . Skowronek, \it Three-by-three bound entanglement with general
unextendible product bases, \rm J. Math. Phys. \bf 52 \rm (2011), 122202.

\bibitem{slm}
P. \O . Sollid, J. M. Leinaas, and J. Myrheim, \it Unextendible
product bases and extremal density matrices with positive partial
transpose, \rm Phys. Rev A \bf 84 \rm (2011), 042325.

\bibitem{stormer}
E. St\o rmer,
\it Positive linear maps of operator algebras,
\rm Acta Math. \bf 110 \rm (1963), 233--278.

\bibitem{stormer82}
E. St\o rmer, \it Decomposable positive maps on $C^*$-algebras,
\rm Proc. Amer. Math. Soc. \bf 86 \rm (1982), 402--404.

\bibitem{wall}
N. R. Wallach,
\it An Unentangled Gleason's Theorem,
\rm Contemp. Math. \bf 305 \rm (2002), 291--298.

\bibitem{woronowicz}
S. L. Woronowicz, \it Positive maps of low dimensional matrix algebras,
\rm Rep. Math. Phys. \bf 10 \rm (1976), 165--183.


\end{thebibliography}
\end{document}